\documentstyle[12pt]{article}

\def\thefigurelist#1{\section*{Figure Captions}

\list {[\arabic{figure}]}{\settowidth\labelwidth{[#1]}\leftmargin\labelwidth
\advance\leftmargin\labelsep

 \usecounter{figure}}

 \def\newblock{\hskip .11em plus .33em minus .07em}

 \sloppy\clubpenalty4000\widowpenalty4000

   \sfcode`\.=1000\relax}

\font\twelve=cmbx10 at 15pt
\font\ten=cmbx10 at 12pt

\begin{document}

\begin{titlepage}

\begin{center}

{\ten Centre de Physique Th\'eorique - CNRS - Luminy, Case 907}

{\ten F-13288 Marseille Cedex 9 - France }

{\ten Unit\'e Propre de Recherche 7061}

\vspace{1 cm}

{\twelve A STOCHASTIC PROCESS}

{\twelve FOR THE DYNAMICS}

{\twelve OF THE TURBULENT CASCADE}

\vspace{0.3 cm}

{\bf R. LIMA and R. VILELA MENDES}\footnote{Centro de F\'{i}sica
da Mat\'{e}ria Condensada, Av. Gama Pinto 2, P-1699 Lisboa Codex,
Portugal}

\vspace{0,6 cm}

{\bf Abstract}

\end{center}

Velocity increments over a distance r and turbulent energy
dissipation on a box of size r are well described by the
 multifractal models of fully developed turbulence. These
quantities and models however, do not involve time-correlations and
therefore are not a detailed test of the dynamics of the
turbulent cascade.

If the time development of the turbulent cascade, in the inertial
range, is related to the lifetime of the eddies at different lenght
scales, the time correlations may be described by a stochastic
process on a tree with jumping kernels which are a function of the
ultrametric (tree) distance. We obtain the solutions of the
Chapman-Kolmogorov equation for such a stochastic process, with
jumping kernels depending on the ultrametric distance, but with an
arbitrarily specified invariant probability measure.
We then show how to use these solutions to compute the time
correlations in the turbulent cascade.

\vspace{0,8 cm}

\noindent Number of figures : 2

\bigskip

\noindent November 1993

\noindent CPT-93/P.2965

\bigskip

\noindent anonymous ftp or gopher: cpt.univ-mrs.fr

\end{titlepage}

\section {Introduction}

One of the most interesting phenomena in fully developed turbulence is
the occurence of an energy cascade from the macroscopic length scale L
of the experimental apparatus down to smaller and smaller length scales.
Lenght scales $l$ in the range $L>>l>>\eta $, $\eta$ being the scale where
the fragmentation process is stopped by dissipation, are said to be in
the "inertial range". In the inertial range viscosity effects are not
important and Kolmogorov$^{1}$  proposed long ago a scaling theory
with conserved energy transfer between length scales. From scale
invariance and the assumption that turbulence is space-filling it
follows that the velocity fluctuation $\delta v(l)$ over an active eddy of
size $l$ scales as

$$
\langle \|\delta v(l)\|^{p}\rangle \sim l^{\zeta _{p}} \eqno\hbox{(1.1)}
$$
with $\zeta =\frac {p}{3}$.

\noindent However turbulence may or may not be space-filling$^{2}$ and the
volume of the active eddies may change when the energy is transferred
{}from the scale $l_{n}$ to the scale $l_{n+1}$. This leads naturally
to a fractal
structure for the cascade with fractal dimension less than 3. For
example in the
$\beta$-model$^{3}$ the rate of energy transfer
$$E_{n} \sim
\frac {\delta v_{n}^{3}}{l_{n}}$$
does not change along the cascade but the total mass of the active eddies is
multiplied by $\beta $ at each step. Then the exponent $\zeta _{p}$ in Eq.(
1.1) becomes
$$
 \zeta _{p}=p.h+3-D
$$
with $h=\frac {D-2}{3}$, where $D$ is the fractal dimension related to $
\beta $ by
$$
  \log_{2}\beta =D-3
$$
if the length scales are related by $l_{n}=l_{0} 2^{-n}$.

\noindent The $\beta$-model as well as a log-normal model$^{4}$ for the
distribution of $E_{n}$ are however in contradiction with the experimental
results on moments of higher order for the velocity structure functions$^
{5,6}$. This fact led to the proposal of a multifractal
generalization$^{7,8}$ called the random $\beta$-model where it is
assumed that, at each scale $l_{n}$, there are several distinct
$\beta _{n}(k)$'s
which are chosen according to some probability law. That is, the energy
transfer may take place according to several distinct dimensional
routes. Requiring a fixed energy transfer rate one obtains
$$
  \frac {\delta v_{n}^{3}(k)}{l_{n}}=\beta _{n+1}(k)\frac {\delta v_{n+1}
^{3}(k)}{l_{n+1}} \eqno\hbox{(1.2)}
$$
Hence at scale $l_{n}$ the velocity fluctuation in each eddy depends on the
fragmentation history which is defined by the product $\beta _{1}.\beta
_{2}.^{...}.\beta _{n}$.
Then
$$
  \delta v_{n} \sim l_{n}^{\frac {1}{3}} \bigl(\prod_{i=1}^{n} \beta _{i}
\bigr)^{-\frac {1}{3}}  \eqno\hbox{(1.3)}
$$
and
$$
  \langle \|\delta v_{n}(l_{n})\|^{p}\rangle \sim l_{n}^{\frac {p}{3}} \int
\prod_{i=1}^{n}d\beta _{i} \beta _{i}^{1-\frac {p}{3}}
  P(\beta _{1},^{...},\beta _{n}) \eqno\hbox{(1.4)}
$$
P($\beta _{1},^{...},\beta_{n}$) being the occurence probability of the
sequence
$\beta_{1},^{...},\beta_{n}$.

\noindent At the level of precision of the existing experiments, agreement
with the data is already obtained if one assumes independent fragmentations
$$
  P(\beta _{1},^{...},\beta _{n})=\prod_{i=1}^{n} P(\beta _{i}) \eqno\hbox{
(1.5a)}
$$
and a simple binomial process
$$
  P(\beta)=\gamma  \delta _{1} + (1-\gamma ) \delta _{\frac {1}{2}}
  \eqno\hbox{(1.5b)}
$$
$\gamma$  is a parameter chosen to fit the data ($\gamma \simeq 0.875$).

\noindent The assumptions (1.5) only define the probability distribution
$\rho_{i}$ at each level of the cascade tree (Fig.1). They make no statement
concerning the time evolution and the time scales of the eddies in the
cascade. In fact it is well known$^{6,9}$ that only a few restrictions
are imposed on the form of the velocity fields by the predictions of
the statistical models described above. These are essentially the
existence of singularities in the derivatives of the velocity field
at some points.
Aside from that, several distinct velocity fields may be compatible
with the spectra and the scaling laws. They range from the superposition
of random uncorrelated Gaussian components having only a correct spectrum
in the inertial range$^{10,11}$ to simple flow fields in isolation, such
as a vortex sheet wrapping up while being stretched by a large-scale
straining motion$^{12}$.

\noindent One way to further the research on the dynamical properties of the
velocity field is to study time-correlations for the observables.
To make a connection of the models with the time dependence of each
observable one should first consider the dynamical aspects of the
energy cascade itself. To do this a time hierarchy in the development
of the turbulent cascade must be defined. In what follows we deal with
this issue.

\noindent For a binary cascade tree (Fig.1) we may use a dyadic labelling for
the possible states at each level. The {\it state space} $V_{n}$ at level
$l_{n}$ is the set of all products $\beta^{(1)}.^{...}.\beta^{(n)}$ with
 $\beta ^{(i)}\in {\beta_{0},\beta_{1}}$. (In
Eq.(1.5b) $\beta_{0}=1$ and $\beta_{1}=\frac {1}{2}$). There are $2^{n}$
elements in $V_{n}$ and the
probability of the state i is
$$
  \rho _{i} = P(\beta _{}^{(1)}.^{...}.\beta ^{(n)}) = \gamma ^{n_{0}(i)}
(
1-\gamma )^{n_{1}(i)} \eqno\hbox{(1.6)}
$$
where $n_{0}$(i) and $n_{1}$(i) are the number of zeros and ones in the
dyadic
labelling of the state i.

\noindent From the random $\beta$-model all one obtains is a statement about
these probabilities.  This suffices to interpret most of the current
experimental results which concern mostly velocity increments over a
distance r and turbulent energy dissipation over a box of size r.
These quantities do not involve time-correlations and therefore do not
make a detailed test of the dynamics of the cascade, they only test its
invariant probability measure.
As discussed above to identify the physical mechanisms
behind the structure of fully developed turbulence, more information is
needed. If one wants, for example, the time
correlations at a point moving with the free-stream velocity of the
fluid one should explicitly consider models for the dynamics
in state space at each level n.
To the same invariant measure ${\rho _{i}}$ correspond many different
processes. The most unstructured process corresponds to the statement
that, if at time zero one finds the state i, then the transition
probability to the state j at time t is proportional to $\rho_{j}$.
For the turbulent cascade the unstructured process does not seem to be
natural because, if the lifetime of the eddies in the inertial range
scales like $\frac {l_{n}}{\delta v_{n}}$, then we expect larger eddies to
live longer than small
eddies. That is, if at time t the fluctuation $\delta v_{n}(x)$ at the
point $x$
is receiving its energy through a fragmentation history leading to the
state i then, a short time thereafter, we expect to find a different
state which is nearby in the sense of the natural ultrametric distance
in the tree.

\noindent To characterize a stochastic process on a tree one has to solve the
Chapman-Kolmogorov equation for the transition probabilities
$$
  \partial_{t}p(zt|y0)=\int dx \{W(z|xt) p(xt|y0) - W(x|zt) p(zt|y0)\}
  \eqno\hbox{(1.7)}
$$
with kernels $W(z|xt)$ that reflect the (natural) ultrametric distance in
the tree. For kernels that depend only on the distance $W(z|xt) =
W(|z-x|)$,
Ogielski and Stein$^{13}$ found the solution of Eq.(1.7).
Albeverio and Karwowski$^{14,15}$ have also constructed the
stochastic processes on arbitrary p-adic fields $Q_{p}$ for the case where
the jumping kernels depend only on the distance between p-adic balls
(see also Brekke and Olson$^{16}$). However it is easy to see from
the equation for the probability densities
$$
  \partial_{t}\rho (zt)=\int dx \{W(z|xt) \rho (xt) - W(x|zt) \rho (zt)\}
  \eqno\hbox{(1.8)}
$$
that if $W(z|xt)=W(|z-x|)$ then the invariant density is $\rho$(z)=const.
For the stochastic process of the turbulent cascade we require a
non-constant invariant density as in Eq.(1.6) and the results of the
authors of Refs. 13-16 cannot be used.

\noindent From (1.8) if follows that with
$$
  W(z|xt) = \rho (z) f(|z-x|) \eqno\hbox{(1.9)}
$$
the invariant density is $\rho$(z) and, at the same time, full account is
taken of the dependence of the transition probability on the distance
between the points z and x in state space. In the next Section we
characterize the solutions of the Chapman-Kolmogorov equation for
kernels of the form (1.9). In Section 3 we then show how to use these
solutions to compute (or parametrize) the time correlations of the
turbulent cascade.

\section{Random walk on a tree with asymmetric
jumping kernels}

We rewrite Eq.(1.8) in matrix form
$$
  \frac {\partial}{\partial t}\rho (t) = W \rho (t)  \eqno\hbox{(2.1)}
$$
where W is the matrix
$$
 \left( \matrix{W_{11} & \rho_{1}\epsilon_{1} & \rho_{1}\epsilon_{2} &
\rho_{1}\epsilon_{2} & \rho_{1}\epsilon_{3} & \rho_{1}\epsilon_{3} &
 \rho_{1}\epsilon_{3} & \rho_{1}\epsilon_{3} &
  . & . & . \cr
  \rho_{2}\epsilon_{1} & W_{22} & \rho_{2}\epsilon_{2} & \rho_{2}
\epsilon_{2} & \rho_{2}\epsilon_{3} & \rho_{2}\epsilon_{3} &
\rho_{2}\epsilon_{3} & \rho_{2}\epsilon_{3} &
  . & . & . \cr
  \rho_{3}\epsilon_{2} & \rho_{3}\epsilon_{2} & W_{33} &
\rho_{3}\epsilon_{1} & \rho_{3}\epsilon_{3} & \rho_{3}\epsilon_{3} &
\rho_{3}\epsilon_{3} & \rho_{3}\epsilon _{3} &
  . & . & . \cr
  \rho _{4}\epsilon _{2} & \rho _{4}\epsilon _{2} & \rho _{4}\epsilon
_{1} & W_{44} & \rho _{4}\epsilon _{3} & \rho _{4}\epsilon _{3} &
\rho _{4}\epsilon _{3} & \rho _{4}\epsilon _{3} &
  . & . & . \cr
  \rho _{5}\epsilon _{3} & \rho _{5}\epsilon _{3} & \rho _{5}\epsilon _{3} &
\rho _{5}\epsilon _{3} & W_{55} & \rho _{5}\epsilon _{1} & \rho _{5}
\epsilon _{2} & \rho _{5}\epsilon _{2} &
  . & . & . \cr
  \rho _{6}\epsilon _{3} & \rho _{6}\epsilon _{3} & \rho _{6}\epsilon _{3} &
\rho _{6}\epsilon _{3} & \rho _{2}\epsilon _{1} & W_{66} & \rho _{6}
\epsilon _{2} & \rho _{6}\epsilon _{2} &
  . & . & . \cr
  \rho _{7}\epsilon _{3} & \rho _{7}\epsilon _{3} & \rho _{7}\epsilon _{3} &
\rho _{7}\epsilon _{3} & \rho _{7}\epsilon _{2} & \rho _{7}\epsilon _{2} &
W_{77} & \rho _{7}\epsilon _{1} &
  . & . & . \cr
  \rho _{8}\epsilon _{3} & \rho _{8}\epsilon _{3} & \rho _{8}\epsilon _{3} &
\rho _{8}\epsilon _{3} & \rho _{8}\epsilon _{2} & \rho _{8}\epsilon _{2} &
\rho _{8}\epsilon _{1} & W_{88} &
  . & . & . \cr
    .  &  .  &  .  &  .  &  .  &  .  &  .  &  .  &
  . & . & . \cr
    .  &  .  &  .  &  .  &  .  &  .  &  .  &  .  &
  . & . & . \cr}\right)
$$
Notice that the matrix has increasingly larger non-diagonal blocks of
size $2^{i}\times 2^{i}$ which have the common factor $\epsilon_{i}$. These
blocks
correspond to jumps to a ultrametric distance i. A matrix element in a
block of size $2^{i}\times 2^{i}$ at line k equals $\rho _{k}\epsilon _{i
}$, $\epsilon _{i}$ being the value
of $f(|z-x|)$ in Eq.(1.9) for a jump to a distance i. The elements
$W_{ii}$ in the
diagonal are such that the columns add to zero.

\noindent As in the symmetric case studied by Ogielski and Stein$^{13}$ we
find the complete set of eigenvectors of the matrix W. For a matrix of
dimension $2^{n}$, which describes the stochastic process at the nth level of
the tree, the eigenvectors are:

\noindent (i) The eigenvector
$$
\left( \matrix{\rho _{1} \cr \rho _{2} \cr . \cr . \cr . \cr \rho _{2^{n
}} \cr}\right)
$$
with eigenvalue $\lambda _{0}=0$;

\noindent (ii) n classes with $2^{n-k}$ (k=1 ... n) eigenvectors each, where
each eigenvector has only $2^{k}$ non-zero elements. The first $2^{k-1}$
elements are
positive and the others are negative. The $2^{k}$ non-zero elements of an
eigenvector have a common ancestor, in the tree, at the level n-k. The
non-zero elements of an (unnormalized) eigenvector are formed by
multiplying the corresponding $\rho_{i}$ by the sum of the $\rho_{j}$'s of
the complementary group in the non-zero set of elements. The formation rule
is easier to understand from an example.

\noindent Let n=3 (Fig. 2). There are then three classes of eigenvectors in
the group (ii), typical examples of which are:

\noindent a)
$$
\left(\matrix{\rho _{1}\rho _{2} \cr -\rho _{2}\rho _{1} \cr 0 \cr 0 \cr
 0 \cr 0 \cr 0 \cr 0 \cr }\right)$$
 $\lambda _{1}=-(\epsilon _{1}(\rho _{1}+\rho _{2})+\epsilon _{2}(\rho _
{3}+\rho _{4})+\epsilon _{3}(\rho _{5}+\rho _{6}+\rho _{7}+\rho _{8}))
$

\noindent Four eigenvectors of this type corresponding to the independent
groups of two elements with a common ancestor at level 2.

\noindent b)
$$
\left(\matrix{\rho _{1}(\rho _{3}+\rho _{4}) \cr \rho _{2}(\rho _{3}+
\rho _{4})_{} \cr -\rho _{3}(\rho _{1}+\rho _{2}) \cr -\rho _{4}(\rho _{1
}+\rho _{2})
  \cr 0 \cr 0 \cr 0 \cr 0 \cr}\right)$$
$\lambda _{5}=-(\epsilon _{2}(\rho _{1}+\rho _{2}+\rho _{3}+\rho _{4})+
\epsilon _{3}(\rho _{5}+\rho _{6}+\rho _{7}+\rho _{8}))
$

\noindent Two eigenvectors of this type.

\noindent c)
$$
\left(\matrix{\rho _{1}(\rho _{5}+\rho _{6}+\rho _{7}+\rho _{8}) \cr
\rho _{2}(\rho _{5}+\rho _{6}+\rho _{7}+\rho _{8})_{} \cr \rho _{3}(\rho
_{5}+\rho _{6}+\rho _{7}+\rho _{8}) \cr
  \rho _{4}(\rho _{5}+\rho _{6}+\rho _{7}+\rho _{8}) \cr -\rho _{5}(\rho
_{1}+\rho _{2}+\rho _{3}+\rho _{4}) \cr -\rho _{6}(\rho _{1}+\rho _{2}+
\rho _{3}+\rho _{4}) \cr -\rho _{7}(\rho _{1}+\rho _{2}+\rho _{3}+\rho
_{4 })
  \cr -\rho _{8}(\rho _{1}+\rho _{2}+\rho _{3}+\rho _{4}) \cr}\right)$$
$\lambda _{7}=-\epsilon _{3}(\rho _{1}+\rho _{2}+\rho _{3}+\rho _{4}+
\rho _{5}+\rho _{6}+\rho _{7}+\rho _{8})=-\epsilon _{3}
$

\noindent One eigenvector of this type.

\noindent The rule of formation for the eigenvalues is clear from the example
above. Each eigenvalue is a sum of terms
$$
  \lambda _{i}=- \sum_{j=l(i)}^{n} \epsilon _{j} {\sum_{k} \rho _{k}}
\eqno
\hbox{(2.2)}
$$
where $\epsilon _{l(i)}$ is the first $\epsilon$  which covers all the
non-zero
 elements of the
vector and the sum $\sum_{k} \rho _{k}$ contains all the probability
densities of
the states reached by an $\epsilon_{j}$-jump.

\noindent A solution of Eq.(2.1) is an arbitrary superposition
$$
  \rho (t)=\sum c_{i} e^{-\lambda _{i}t} v_{i}  \eqno\hbox{(2.3)}
$$
of the eigenvectors above. From (2.3) it is now easy to construct the
general solution of the Chapman-Kolmogorov equation. To obtain the
transition probability $p(j t|i 0)$ from the state i at time zero to the
state j at time t, one chooses the coefficients $c_{i}$ in (2.3) in such a
way that, at time zero, only the i-th component is non-zero and then
read the value of the j-th component at time t. Before stating the
general result we illustrate it by writing the transition probabilities
for transitions between typical states in each group for the case n=3.
$$
  p(1t|10)=\frac {\rho _{1}}{\rho _{1}+^{...}+\rho _{8}} + \frac {\rho _{1
}\rho _{2}}{\rho _{1}(\rho _{1}+\rho _{2})}
  e^{-t\{\overbrace{\epsilon _{1}(\rho _{1}+\rho _{2})+\epsilon _{2}(
\rho _{3}+\rho _{4})+\epsilon _{3}(\rho _{5}+^{...}+\rho _{8})}^{\lambda
_{1}}\}}$$

$$ + \frac {\rho _{1}(\rho _{3}+\rho _{4})}{(\rho _{1}+\rho _{2})(
\rho _{1}+\rho _{2}+\rho _{3}+\rho _{4})}
  e^{-t\{\overbrace{\epsilon _{2}(\rho _{1}+^{...}+\rho _{4})+\epsilon _{3
}(\rho _{5}+^{...}+\rho _{8})}^{\lambda _{5}}\}}$$

$$ +  \frac {\rho _{1}(\rho _{5}+^{...}+\rho _{8})}{(\rho _{1}+^{...}+
\rho _{4})(\rho _{1}+^{...}_{}+\rho _{8})} e^{-t\epsilon _{3}}$$

$$ p(2t|10)=\frac {\rho _{2}}{\rho _{1}+^{...}+\rho _{8}} - \frac {\rho _{2
}\rho _{1}}{\rho _{1}(\rho _{1}+\rho _{2})} e^{-t\lambda _{1}} +  \frac {
\rho _{2}(\rho _{3}+\rho _{4})}{(\rho _{1}+\rho _{2})(\rho _{1}+\rho _{2}+
\rho _{3}+\rho _{4})} e^{-t\lambda _{5}}$$

$$ + \frac {\rho _{2}(\rho _{5}+^{...}+\rho _{8})}{(\rho _{1}+^{...}+
\rho _{4})(\rho _{1}+^{...}_{}+\rho _{8})} e^{-t\lambda _{7}}$$

$$p(3t|10)=\frac {\rho_{3}}{\rho _{1}+^{...}+\rho _{8}} - \frac {\rho _{3
}(\rho _{1}+\rho _{2})}{(\rho _{1}+\rho _{2})(\rho _{1}+\rho _{2}+\rho _{3
}+\rho _{4})} e^{-t\lambda _{5}}
  +  \frac {\rho _{3}(\rho _{5}+^{...}+\rho _{8})}{(\rho _{1}+^{...}+
\rho _{4})(\rho _{1}+^{...}_{}+\rho _{8})} e^{-t\lambda _{7}}$$

$$p(5t|10)=\frac {\rho _{5}}{\rho _{1}+^{...}+\rho _{8}} - \frac {\rho _{5
}(\rho _{1}+^{...}+\rho _{4})}{(\rho _{1}+^{...}+\rho _{4})(\rho _{1}+^{...
}_{}+\rho _{8})} e^{-t\lambda _{7}}    \eqno\hbox{(2.4)}
$$
Of course in this case $\rho_{1}+^{...}+\rho _{8}=1$ but we have kept this
term to
emphasize the rule of formation of the coefficients. The general rule
for the transition probability $p(j t|i 0)$ between two states i and j at
the level n in the tree is the following:

\noindent (i) $p(j t|i 0)$ is a sum of terms, the first of which is $\rho_{j}$
(the target
probability), and has as many terms as the number of eigenvectors that
have non-zero elements in both the i and the j positions.
$$
  p(j t|i 0) = \rho _{j} + \sum_{k} c_{k} e^{-t\lambda _{k}}
\eqno\hbox{(2.5)}
$$

\noindent (ii) The exponential factor in each term contains the eigenvalue of
the associated eigenvector.

\noindent (iii) The coefficients all contain in the numerator the target
probability $\rho_{j}$ multiplied by the sum of the probabilities of non-zero
elements of the corresponding eigenvectors in the half that does not
contain j. The denominator is the sum of the half that contains i
multiplied by the sum of all probabilities associated to the non-zero
elements of the eigenvector.

\noindent (iv) The sign of the coefficient is the product of the signs of the
i and j entries in the eigenvector.

\section{Time correlations. Application to the turbulent cascade}

Once the transition probabilities $p(zt|y0)$, solutions of the
Chapman-Kolmogorov equation are known, the time correlations of the
process are obtained from
$$
  \langle x(t) x(0) \rangle = \int dy dx \ y \ p(yt|x0)
\ x \
\rho (x) \eqno\hbox{(3.1)}
$$
or
$$
  \langle x(t) x(0) \rangle = \sum_{i,j} x_{j} p(jt|i0) x_{i} \rho _{i} \eqno
\hbox{(3.2)}
$$
for a discrete state space.

\noindent Using the results of Section 2 (Eq.(2.5)) one sees that at large
times  the time
correlation at level n will be dominated by the largest non-zero
eigenvalue $\lambda _{2^{n}-1}=-\epsilon _{n}$. Assuming that the dynamics
of the turbulent
cascade is controlled by the decay of the eddies, the largest non-zero
eigenvalue will always be the same, associated to the mean lifetime of
large eddies. However the asymptotic long-time correlation will be
difficult to measure because of the small values of $\langle x(t) x(0)
\rangle$ at large t. Error bars, in numerical or actual experiments,
are likely to be larger than $e^{-\epsilon_{n}t}$ for t large.

\noindent If, as we are proposing, the time correlations in the
turbulent cascade are described by a stochastic process with kernels
that depend on tree distances, a first qualitative
prediction is the occurence of several exponential slopes, as the time
increases, in the time-correlation functions. Notice that the existence
of different time scales, as a consequence of the advection of
small-scale eddies by large-scale motions, was already pointed out by
Kolmogorov (see Ref.9).

\noindent Of special interest is the slope of the short-time correlation
which is controlled by the smallest eigenvalue. Using the dyadic expansion to
label the points $x_{i}$ in state space
$$
  x_{i} = \beta ^{n_{0}(i)}_{0} \beta ^{n_{1}(i)}_{1} \eqno\hbox{(3.3)}
$$

$$
  \rho _{i} = \gamma ^{n_{0}(i)} (1-\gamma )^{n_{1}(i)} \eqno\hbox{(3.4)}
$$
where $n_{0}$(i) and $n_{1}$(i) are the number of zeros and ones in the
dyadic expansion of i. Assuming $\gamma >(1-\gamma)$ the smallest eigenvalue
for the
dynamics at level n is
$$
  \lambda _{1}^{(n)} = -
  \{\epsilon _{1}^{(n)}(\rho _{1}^{(n)}+\rho _{2}^{(n)})+\epsilon _{2}^{(n)
}(\rho _{3}^{(n)}+\rho _{4}^{(n)})+\epsilon _{3}^{(n)}(\rho _{5}^{(n)}+^{..
{}.}+\rho _{8}^{(n)})
  +\epsilon _{4}^{(n)}(\rho _{9}^{(n)}+^{...}+\rho _{16}^{(n)})+^{.....}\}
\eqno\hbox{(3.5)}
$$

\noindent If the dynamics of the turbulent cascade is associated to the decay
of the eddies of different sizes, it is reasonable to assume that
$$
  \epsilon _{i^{}}^{(n-1)} = \epsilon _{i+1}^{(n)}  \eqno\hbox{(3.6)}
$$

Using this relation and the relations between the probability densities
at the levels n and n-1 one obtains
$$
  \lambda _{1}^{(n)} - \lambda _{1}^{(n-1)} = -(\epsilon _{1}^{(n)} -
\epsilon _{1}^{(n-1)}) \rho _{1}^{(n-1)} \eqno\hbox{(3.7)}
$$

\noindent One concludes that the ratio of short-time correlations measures the
difference between the lifetimes of the structures at different length
scales. If the dynamics of the cascade is controlled by the decay of the
eddies and these have different lifetimes at different scales, the
ultrametric stochastic model is an appropriate way to parametrize the
dynamics and to characterize it in quantitative terms. Other models
yield different correlation structures.

\noindent Notice that here we are concerned with the time fluctuations of the
turbulent cascade itself, not with the changes induced by the overall
motion of the fluid. This means that for a fluid in motion with
free-stream velocity $\overrightarrow{U}$ the correlations to measure, for
an observable $\Delta $, are
$$
  \langle \Delta (x+\overrightarrow{U}t,t) \Delta (x,0) \rangle
$$
The measure of the short-time behaviour of such quantities
and the detection of several time scales in the time-correlations
will test the usefulness of the turbulent cascade process proposed
in this paper. Notice however that, in particular, the accuracy needed to
detect different time scales, is a great experimental challenge.

{\bf Figure captions}

Fig.1  The state space at level n for a dyadic turbulent cascade

Fig.2  The three types of stochastic transitions associated
to three different classes of eigenvectors

\end{document}